\documentclass[a4paper]{jpconf}

\usepackage{graphicx}
\usepackage{wasysym}
\usepackage{amssymb}
\usepackage{amsthm}

\newcommand{\OO}{{\cal O}}
\newcommand{\CFT}{{\rm \mbox{{\scriptsize  CFT}}}}
\newcommand{\dil}{{\rm \mbox{{\scriptsize  dil}}}}
\newcommand{\LL}{{\cal L}}
\newcommand{\ddef}{{\rm \mbox{{\scriptsize  def}}}}

\newcommand{\UV}{{\rm \mbox{{\tiny  UV }}}}
\newcommand{\IR}{{\rm \mbox{{\tiny  IR }}}}
\newcommand{\PP}{{\rm \mbox{{\tiny  P}}}}
\newcommand{\NPP}{{\rm \mbox{{\tiny  NP}}}}
\newcommand{\phys}{{\rm \mbox{{\scriptsize  phys}}}}
\newcommand{\ma}{{\rm \mbox{{\scriptsize  M}}}}

\begin{document}

\begin{flushright}
MPP-2015-217
\end{flushright}

\title{A holographic realization of light dilatons}

\author{Eugenio Meg\'{\i}as}

\address{Max-Planck-Institut f\"ur Physik (Werner-Heisenberg-Institut), F\"ohringer Ring 6, D-80805, Munich, Germany}

\ead{emegias@mppmu.mpg.de}

\begin{abstract}
We study a realization of light dilatons by considering Conformal Field Theories (CFT) deformed by a nearly marginal operator~${\cal O}$. This is discussed in the holographic setup consisting of Renormalization Group (RG) flows that end on a soft wall. We obtain a mass formula for the dilaton as an average along the RG flow. Finally we discuss the holographic method to compute the condensate~$\langle {\cal O} \rangle$.
\end{abstract}

\section{Introduction}
\label{sec:intro}

Spontaneous breaking of conformal invariance (SBCI) is nowadays a scenario that is extensively studied in the context of the most fundamental problems in physics. On the one hand, this symmetry breaking may give rise to a Goldstone boson, the {\it dilaton}, which might have important consequences in cosmology and particle physics. In particular, some interesting beyond the Standard Model (SM) realizations assume that the SM can be part of a nearly-conformal sector. The electroweak scale would arise from the SBCI, and a naturally light dilaton would appear in the spectrum~\cite{Coradeschi:2013gda,Megias:2014iwa,Cox:2014zea}. On the other hand, the effective theory for the dilaton presents a toy version of the Cosmological Constant (CC) problem~\cite{Bellazzini:2013fga}. Both applications are connected with the concept of naturalness of physical theories. 

Some significant progress on these issues have been achieved recently thanks to the observation by Contino, Pomarol and Rattazzi (CPR)~\cite{CPR,Rattazzi:2010,Pomarol:2010} that a light dilaton should occur naturally in certain deformations of CFTs.  While the original holographic implementation of the CPR proposal is dual to a CFT that contains at least two scalar operators, in this manuscript we try to test it in slightly simpler and perhaps more realistic models, involving a single nearly marginal operator~$\OO$. Under a relevant deformation one generally expects that the operator develops a condensate~$\langle \OO \rangle$, and the fluctuation of~$\langle \OO \rangle$ can already play the role of a dilaton. This will be tested using the techniques developed in the context of holographic RG flows~\cite{DeWolfe:1999cp,deBoer:1999xf,Anselmi:2000fu,Papadimitriou:2004ap,Papadimitriou:2004rz,Papadimitriou:2007sj} and soft-wall models~\cite{Gubser:2000nd,Karch:2006pv}.

Let us include a brief note on related literature. The interest for the dilaton as the pseudo-Goldstone boson of SBCI has a long story~\cite{Gildener:1976ih}. More recently there has been a renewed interest mostly triggered by LHC phenomenology, see e.g.~\cite{Goldberger:2008zz,Vecchi:2010gj}. The identification of a dilaton in holographic RG flows goes back to~\cite{Bianchi:2001de}, though some of the models involve what in the present manuscript we classify as tuned SBCI~\cite{Hoyos:2013gma}. There is also a considerable amount of literature on string-theoretic embeddings  of walking dynamics and light dilatons, see e.g.~\cite{Elander:2013jqa} and references therein. However, in these scenarios it is hard to disentangle whether the dilaton lightness stems from SBCI or from SUSY. In the present work, we consider non-supersymmetric flows.

\section{Spontaneous breaking of conformal invariance and dilatons}
\label{sec:SBCI_dilatons}

Physics at low energies can be captured by effective field theory methods. We will address in this section some basic features of the effective theory for a dilaton, and discuss the mechanism to get a naturally light dilaton recently proposed in~\cite{CPR}.

\subsection{Spontaneous breaking of conformal invariance}
\label{subsec:SBCI}

\begin{figure}[htb]
\includegraphics[width=75mm]{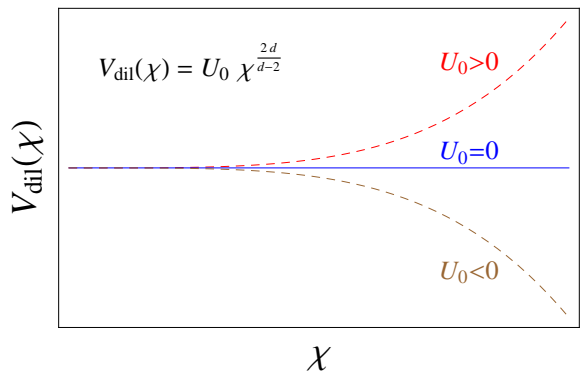} \hspace{2pc}%
\begin{minipage}{18pc}
\vspace{-3.5cm}\caption{Dilaton potential in a CFT for different values of the quartic coupling $U_0$. SBCI is only possible when $U_0=0$. These curves illustrate the results for a potential of the form Eq.~(\ref{eq:VCFT}).}
\label{fig:Vchi}
\end{minipage}
\end{figure}

A scale transformation is defined as $x^\mu \to \frac{1}{b} x^\mu$, where $b$ is the scaling factor. The starting point to compute the effective theory of a dilaton is to construct the most general Lagrangian invariant under this transformation. The scale transformation acts like~$ \chi \to b^{\frac{d-2}{2}} \chi$, in $d>2$ space-time dimensions, where $\chi$ is the canonically normalized dilaton field. Therefore, the  effective Lagrangian compatible with this symmetry allows a potential of the form
\begin{equation}
V^{\CFT}_{\dil}(\chi) = U_0 \, \chi^{\frac{2d}{d-2}} \,. \label{eq:VCFT}
\end{equation}
The value of the ``quartic'' coupling $U_0$, which is the analogue of the CC, is arbitrary. SBCI~means the appearance of a nonzero value for the condensate $\langle \chi \rangle \ne 0$. For $U_0 > 0$ the minimum of the potential is at $\langle \chi \rangle=0$, while for $U_0 < 0$ there is no minimum and the theory is unstable. So, SBCI can only happen in a CFT if $U_0 =0$, and then the dilaton is massless and it is identified as the corresponding Goldstone boson. However, this can only occur if either i) the theory is fine-tuned, or ii) the theory enjoys additional symmetries (SUSY) that enforce $U_0=0$. This picture is illustrated in Fig.~\ref{fig:Vchi}. As we mention in the Introduction, in the following we will consider non-supersymmetric theories. 

\subsection{Naturally light dilatons}
\label{subsec:naturally_light_dilatons}

After the discussion above, a logical question arises: can a naturally light dilaton appear in Quantum Field Theories that are close to conformal invariance (CI)? We will study in the following the simplest situation, consisting of a CFT deformed by a single trace operator $\cal O$, i.e.
\begin{equation}
\LL = \LL^{\CFT} + \delta \LL  \,, \qquad \textrm{with} \qquad \delta \LL = -\lambda \, \OO  \,. \label{eq:Lcftdef}
\end{equation}
The deformation $\delta \LL$ introduces a small explicit breaking of CI if the operator is relevant and nearly marginal, i.e. 
\begin{equation}
{\rm Dim}(\OO) = {d-\Delta} \qquad \textrm{with} \qquad \Delta \ll 1 \,.
\end{equation}
The dimension of the Lagrangian and the coupling constant are  ${\rm Dim}(\LL) = d$ and ${\rm Dim}(\lambda) = \Delta$, respectively. This is the starting point of the CPR mechanism proposed in Refs.~\cite{CPR,Rattazzi:2010,Pomarol:2010}. 

\begin{figure*}[htb]
\begin{tabular}{cc}
\hspace{-0.1cm}\includegraphics[width=75mm]{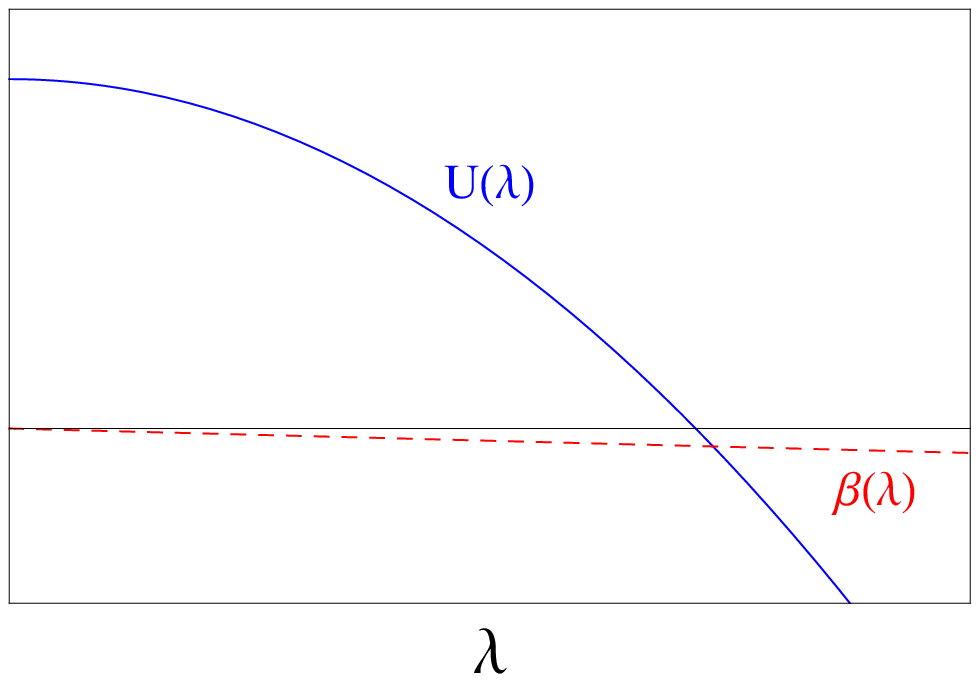} &
\hspace{0.1cm} \includegraphics[width=75mm]{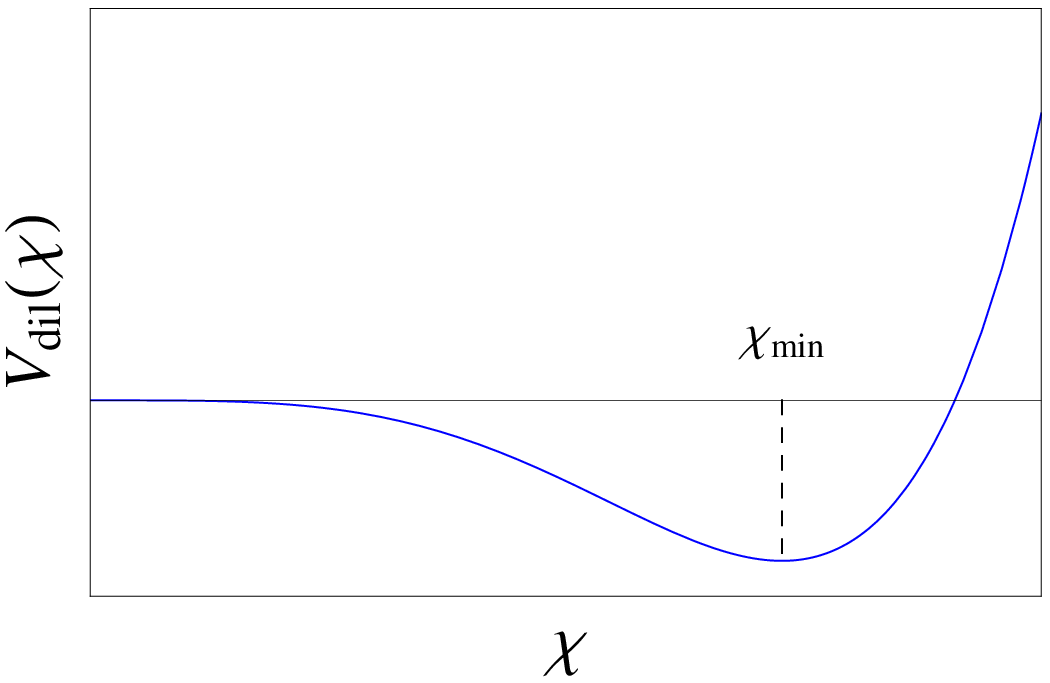} \\
(A)
    &
      (B)
\end{tabular}
\caption{(A) Quartic coupling and beta function as a function of the coupling constant. (B)~Dilaton potential as a function of the dilaton field. These curves illustrate the results for a potential of the form Eq.~(\ref{eq:VCFTdef}).}
\label{fig:Vmin}
\end{figure*}

Let us study now whether SBCI can occur in addition to the explicit breaking of CI. RG~computations capture the $\chi$ dependence of the potential, as the dilaton shifts like a scale transformation, so that the effective potential is obtained by evaluating the running coupling at the scale $\mu = \chi^{\frac{2}{d-2}}$. This generally takes the form
\begin{equation}
V_{\dil}^{\CFT+\ddef}(\chi) =  \chi^{\frac{2d}{d-2}}   U(\lambda(\mu)) \Big|_{\mu=\chi^{\frac{2}{d-2}}}  \,, \label{eq:VCFTdef}
\end{equation}
where $\lambda(\mu)$ is the running coupling constant. The appearance of SBCI amounts to having a nontrivial minimum of the potential. For later convenience, we will refer to this minimum as the infrared (IR) scale~$\chi_\IR$. Since $V_{\dil}^{\CFT+\ddef}{}^\prime(\chi_{\IR}) = \frac{2}{d-2}   \chi_{\IR}^{\frac{d+2}{d-2}} (d \, U(\lambda) + \beta \, U^\prime(\lambda))$, where we have defined the beta function as $\beta = \frac{d\lambda}{d\log\mu}$, it follows that the dilaton mass is small if $\beta$ and $U$ are small at the minimum. Indeed
\begin{equation}
V_{\dil}(\chi_{\IR}) = U_{\textrm{\IR}} \chi_{\IR}^{\frac{2d}{d-2}} \sim \beta_{\IR} \chi_{\IR}^{\frac{2d}{d-2}}  \qquad \textrm{and} \qquad  V_{\dil}^{\prime\prime}(\chi_{\IR}) =  m_{\dil}^2 \sim \beta_{\IR} \chi_{\IR}^{\frac{4}{d-2}}  \,,
\end{equation}
are generically suppressed by one power of $\beta_\IR$. This picture is illustrated in Fig.~\ref{fig:Vmin}. 
In the rest of this manuscript we will present an explicit holographic realization of the CPR proposal.

\section{Holographic description of CFT deformations}
\label{sec:holo_cft}

In this section we review the most important ingredients for the holographic description of CFT deformations. In particular, we stress the importance of the holographic beta function to describe how confinement and condensation of operators arise in the holographic RG~flows.

\subsection{Holographic model}
\label{subsec:holo_model}

Let us consider a dilaton-gravity model in (d+1) dim given by
\begin{equation}
S= {M^{d-1} \over 2} \int d^{d+1}x \sqrt{-g}\left\{  R - (\partial\phi)^2  - 2 V(\phi)\right\} \,.
\end{equation}
Holographically, a CFT deformed by a scalar operator $\OO$ as presented in Eq.~(\ref{eq:Lcftdef}), is realized by a domain wall geometry in which the scalar field $\phi$ develops a profile in the extra dimension $y$,
\begin{equation}
ds^2 = dy^2 + a^2(y)dx^\mu dx^\nu \eta_{\mu\nu} \,, \qquad \phi = \phi(y) \,. 
\end{equation}
The equations of motion for this model can be written as a system of first order equations
\begin{equation}
\frac{d\phi}{dy} = -W^\prime(\phi)   \,, \qquad \frac{1}{a}\frac{da}{dy} = \frac{W(\phi)}{d-1} \,,   \label{eq:eom} 
\end{equation}
where the superpotential $W$ obeys the equation $W^\prime(\phi)^2 = 2 V(\phi) +{d\over d-1} W^2(\phi)$. In a near boundary expansion the scalar field contains two modes,
\begin{equation}
\phi(y) =  \lambda \; e^{-\Delta y} + \langle \OO \rangle \; e^{-(d-\Delta) y} + \cdots  \,, \qquad y \to +\infty \,. \label{eq:phi_expand}
\end{equation}
Following the holographic dictionary, the leading mode is identified with the deformation parameter $\lambda$, while the sub-leading one corresponds to the condensate $\langle \OO \rangle$. These two quantities are eventually related to the two integration constants needed to solve the equations of motion.

\subsection{Holographic beta function}
\label{subsec:holo_beta_function}

\begin{figure*}[htb]
\begin{tabular}{cc}
\includegraphics[width=75mm]{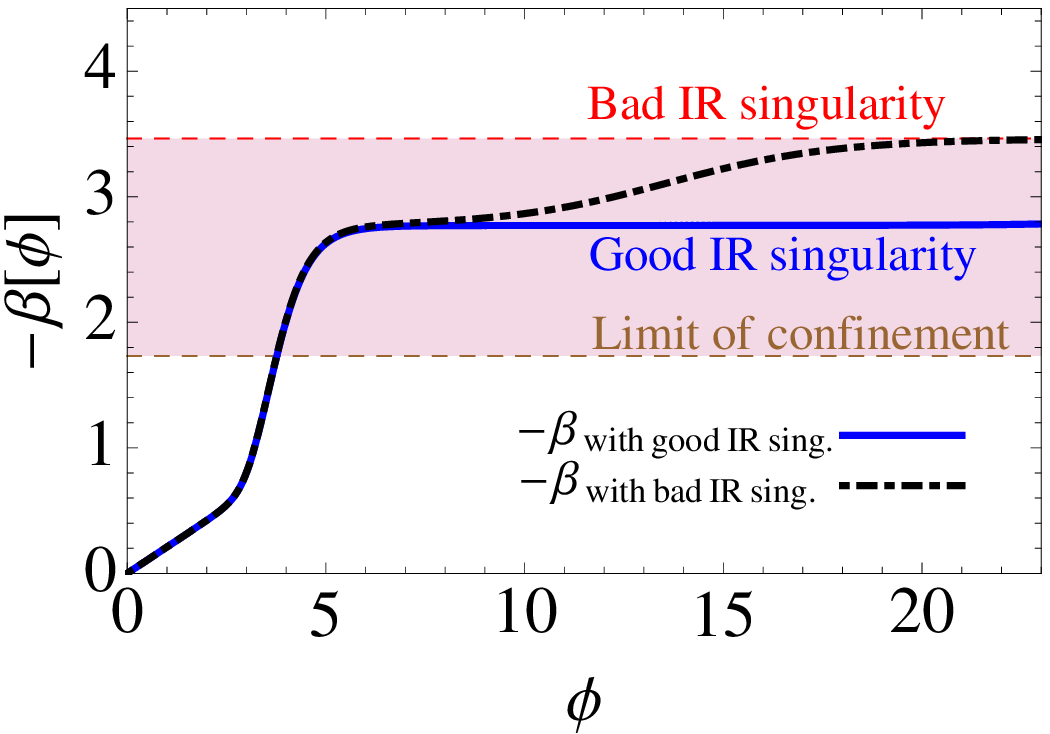} &
\includegraphics[width=75mm]{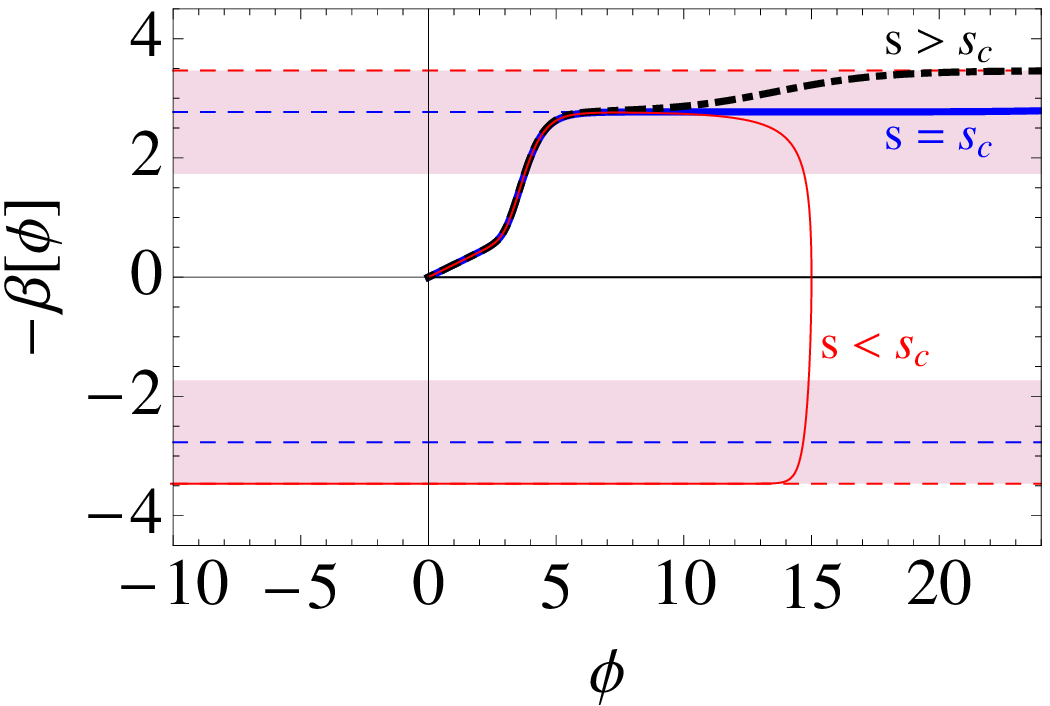} \\
(A)
    &
      (B)
\end{tabular}
\caption{(A) Behavior of $-\beta$ as a function of $\phi$. It is shown two different flows approaching the IR attractors of Eq.~(\ref{eq:attractors}). The shaded band corresponds to the bounds of Eq.~(\ref{eq:betabounds}). (B) RG flow for several values of the integration constant of Eq.~(\ref{eq:betaNP}): $s > s_c$, $s=s_c$ and $s < s_c$. In these plots we have considered $\Delta_-=0.21$.}
\label{fig:betaphi}
\end{figure*}

The warp factor~$a$ plays the role holographically of the renormalization scale~$\mu$. In addition, for a nearly marginal deformation the scalar field close to the boundary behaves as $\phi(y) = \lambda \, e^{-\Delta y} + \cdots  $ with $\Delta \ll 1$, so that it can be approximately identified as the coupling strength of the deformation. Then one immediately identifies the holographic version of the beta function~\cite{deBoer:1999xf,Anselmi:2000fu} as 
\begin{equation}
\beta(\phi) := a \frac{\partial \phi}{\partial a}  = -(d-1) \frac{W^\prime(\phi)}{W(\phi)} \,. \label{eq:beta}
\end{equation}
In the second equality we have made use of the equations of motion Eq.~(\ref{eq:eom}). This definition implies the existence of a 1 to 1 relation between the beta function and the superpotential. It follows also that $\beta$ obeys the following integral equation
\begin{equation}
\left(\beta^2  -{d(d-1) } \right) {W^2 \over 2(d-1)^2} = V  \,.  \label{eq:beta2}
\end{equation}
Since $V(\phi) < 0$, this implies that $\beta$ is bounded everywhere along the flow, i.e. $ |\beta| < \sqrt{d(d-1)}$. Finally, by using Eq.~(\ref{eq:beta2}) and its derivative, one can obtain that $\beta(\phi)$ obeys the first order differential equation
\begin{equation}
\beta\,\beta^\prime = (\beta^2-d(d-1)) \left({\beta\over d-1}+{1\over2}{V^\prime\over V}\right) \,. \label{eq:eqbeta}
\end{equation}
We can easily derive from Eq.~(\ref{eq:eqbeta}) some interesting properties of the beta function. In particular: i) existence of an integration constant, whose physical meaning will be explained in Sec.~\ref{subsec:CFT_confining_def}; ii)~fixed points $\beta=0$ map to extrema of $V$, i.e. $V^\prime = 0$; iii) existence of three IR attractors:  
\begin{equation}
\beta^\prime(\phi)=0 \quad \Longrightarrow \quad \Bigg\{
\begin{tabular}{cc}  
$\beta =  \beta_b \equiv \pm \sqrt{d(d-1)}$  &  \\
\hspace{-0.6cm} $\beta \simeq \beta_g \equiv -\frac{(d-1)}{2}\frac{V^\prime}{V}$  & 
\end{tabular} \,. \label{eq:attractors}
\end{equation}
In the second line of this equation we have assumed that $V^\prime/V$ is smooth. The attractors $\beta = \beta_{b}$ and $\beta = \beta_g$ give rise to singularities of Gubser's bad and good type respectively~\cite{Gubser:2000nd}, provided that $V^\prime/V$ is not too large. The bad (good) attractor is attractive towards large (small) $\phi$. The behavior of $\beta(\phi)$ and the location of the attractors are displayed in Fig.~\ref{fig:betaphi}. Note that $\beta \simeq$ cte implies $V(\phi) \sim e^{\nu\phi}$, so that it is natural to consider an exponential behavior for the scalar potential in the IR. 

\subsection{CFTs with confining deformations}
\label{subsec:CFT_confining_def}

In the following we will concentrate on flows from close to a CFT in the ultraviolet (UV) down to a confining~IR. Some of the main points presented here have been derived previously in e.g.~\cite{Papadimitriou:2007sj,Galow:2009kw,Bourdier:2013axa,Cabrer:2009we}. 

Confinement leads to the generation of a universal mass gap in the spectrum of excitations, and the dilaton is going to be easily identifiable as a low-lying mass-eigenvalue. A simple holographic criterion for confinement is that the conformal coordinate, defined as $dz = -dy/a$, has a finite range. Placing the AdS boundary at $z=0$, the IR singularity is at $z=z_s$ with
\begin{equation}
z_s = \int_0^\infty \frac{d\phi}{a(\phi) W^\prime(\phi)} < \infty \,,
\end{equation}
and the mass gap is identified as $\Lambda_\IR = 1 / z_s$. This leads to the following bounds for the beta function in the IR,
\begin{equation}
\beta(\phi) \to \beta_\infty  \qquad \textrm{with} \qquad  \sqrt{d-1} < -\beta_\infty < \sqrt{d(d-1)} \,.  \label{eq:betabounds}
\end{equation}
The upper bound corresponds to limit of good IR singularity solutions, see Fig.~\ref{fig:betaphi}(A). In the following we will restrict to scalar potentials that are analytic in $\phi$ and even under $\phi \to -\phi$, so that they admit the small $\phi$ expansion
\begin{equation}
V(\phi) = -{d(d-1)\over 2\ell^2} + {M_\phi^2 \over 2} \phi^2 +\OO\big(\phi^4\big) \,.  \label{eq:V}
\end{equation}
The solutions of Eq.~(\ref{eq:eqbeta}) contain generically two contributions that we call perturbative and non-perturbative parts,
\begin{equation}
\beta(\phi) =  \beta_{\PP}(\phi) + \beta_{\NPP}(\phi) \,. \label{eq:betaPNP}
\end{equation}
$\beta_\PP$ is analytic as it is completely fixed by $V$. In particular at first order $\beta_\PP = -\Delta \phi + \dots$, and from Eq.~(\ref{eq:V}) one gets that $\Delta$ is set to either of $\Delta_{\pm}$ with
\begin{equation}
\Delta_\pm = {d\over2}\pm\sqrt{\left({d\over2}\right)^2+M^2_\phi\,\ell^2}  \,.
\end{equation}
Then one finds that there are two types of flows with behavior $\beta_\pm \simeq -\Delta_\pm \phi$. In the case of $\beta_+$ flows, the scalar field behaves as $\phi \simeq 0 + \langle \OO \rangle \, e^{-\Delta_+ y}$. Then one concludes that they correspond to a CFT with SBCI but no explicit breaking. As discussed in Sec.~\ref{sec:SBCI_dilatons}, one expects that this situation is either fine-tuned or supersymmetric. On the other hand, the $\beta_-$ flows lead to the general expansion of Eq.~(\ref{eq:phi_expand}), and so they correspond to a genuine deformation of a CFT with explicit and SBCI. 

$\beta_\NPP$ is a non-analytical function of $\phi^2$ that is controlled by an integration constant. When $\beta_\NPP$ is a small correction to $\beta_\PP$, one can express this solution as~\cite{Papadimitriou:2007sj}
\begin{equation}
\beta_{\NPP}(\phi) \propto s \cdot \exp\left( - \int_0^\phi d\bar\phi \;{d+\beta_\PP^\prime(\bar\phi) \over \beta_\PP(\bar\phi)} \right) \sim s \cdot \phi^{\frac{d}{\Delta_{\pm}}-1}  \,, \label{eq:betaNP}
\end{equation}
where $s$ is the integration constant. For $\beta_+$ flows the only choice that does not break the AdS asymptotic behavior near the boundary is $s=0$. The fine-tuned feature of this kind of flows is then evident from the particular value of the integration constant. On the other hand, $\beta_-$ flows can have in principle any value of $s$, as $\phi^{\frac{d}{\Delta_-}-1}$ is sub-leading to $\phi$.

In order to discuss the physical meaning of the integration constant $s$, let us compute the near boundary expansion of the scalar field for the flow $\beta_-(\phi)= -\Delta_- \phi +\dots -  s \,\phi^{{d\over \Delta_-}-1} +\dots$. From Eqs.~(\ref{eq:eom}) and (\ref{eq:beta}) one gets
\begin{equation}
\phi(y) = \lambda \,e^{-\Delta_- y}(1+\dots) + s\;{\lambda^{{\Delta_+\over \Delta_-}} \over \Delta_+-\Delta_-}  \;\,e^{-\Delta_+ y} \;(1+\dots) \,.
\end{equation}
By comparison with Eq.~(\ref{eq:phi_expand}) we conclude that $s$ is none other than the value of the condensate, i.e. $s \sim \langle \OO \rangle$. In order to determine $\langle \OO \rangle$, the main idea is that $s$ should be adjusted in the UV so that the IR part of the flow is as regular as possible. This selects uniquely a critical value for the integration constant $s=s_c$ which maps to the physical value of the condensate, and this corresponds to a flow falling in the good IR singularity of Eq.~(\ref{eq:attractors}), i.e. $\beta \to -\frac{(d-1)}{2} \frac{V^\prime}{V}$. In effect, as it is shown in Fig.~\ref{fig:betaphi}(B), if one over-shoots (under-shoots) from the UV by taking $s > s_c$ ($s< s_c$), then the solution falls on the bad attractors $\beta \to \mp \sqrt{d(d-1)}$.  Then we conclude that there are always explicit and SBCI flows with $\langle \OO  \rangle \ne 0$. The procedure to find $s_c$ numerically can be made systematic and efficient by shooting instead from the IR, since then the bad attractor is in fact repulsive.

Excitations around the regular flow determines the dilaton mass. We will shortly obtain that it is positive in a large class of models, i.e.~$m_{\dil}^2 > 0$, thus fulfilling the {\it Minimum Energy Theorems} of scalar-gravity systems dual to a CFT with a scalar operator~\cite{Amsel:2006uf,Faulkner:2010fh,Elder:2014fea}. Then, the flow $s_c \sim \langle \OO \rangle_{\phys}$ corresponds to the ground state of the deformed~CFT.

\subsection{Models}

\begin{figure*}[htb]
\begin{tabular}{cc}
\includegraphics[width=75mm]{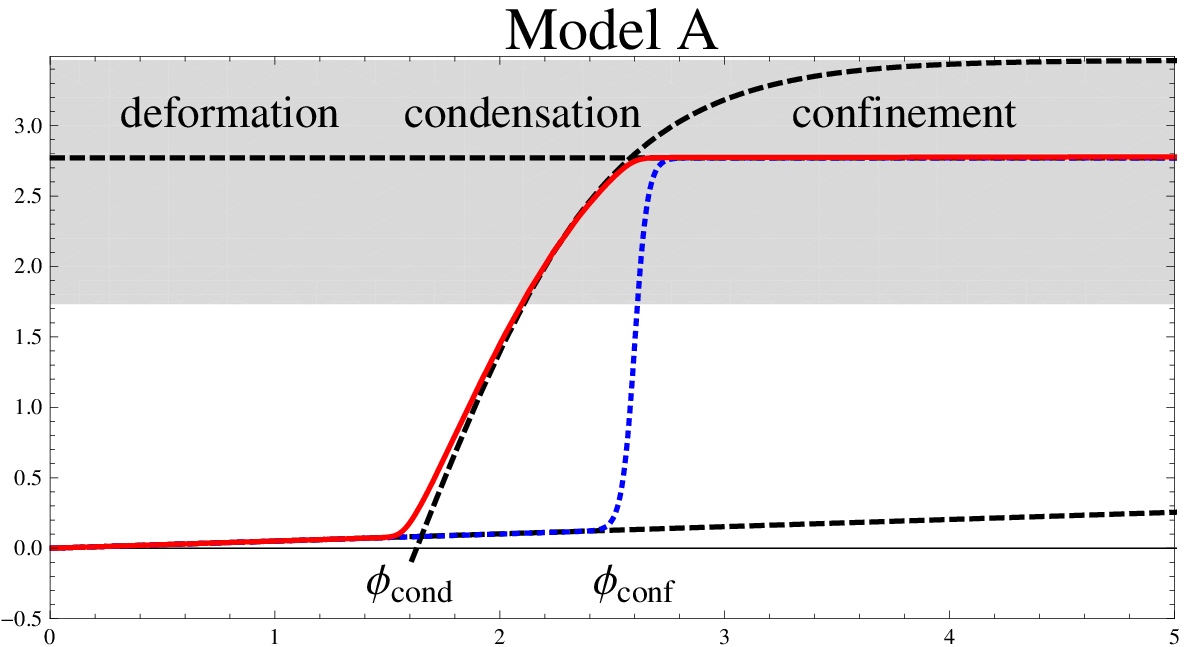} & 
\includegraphics[width=75mm]{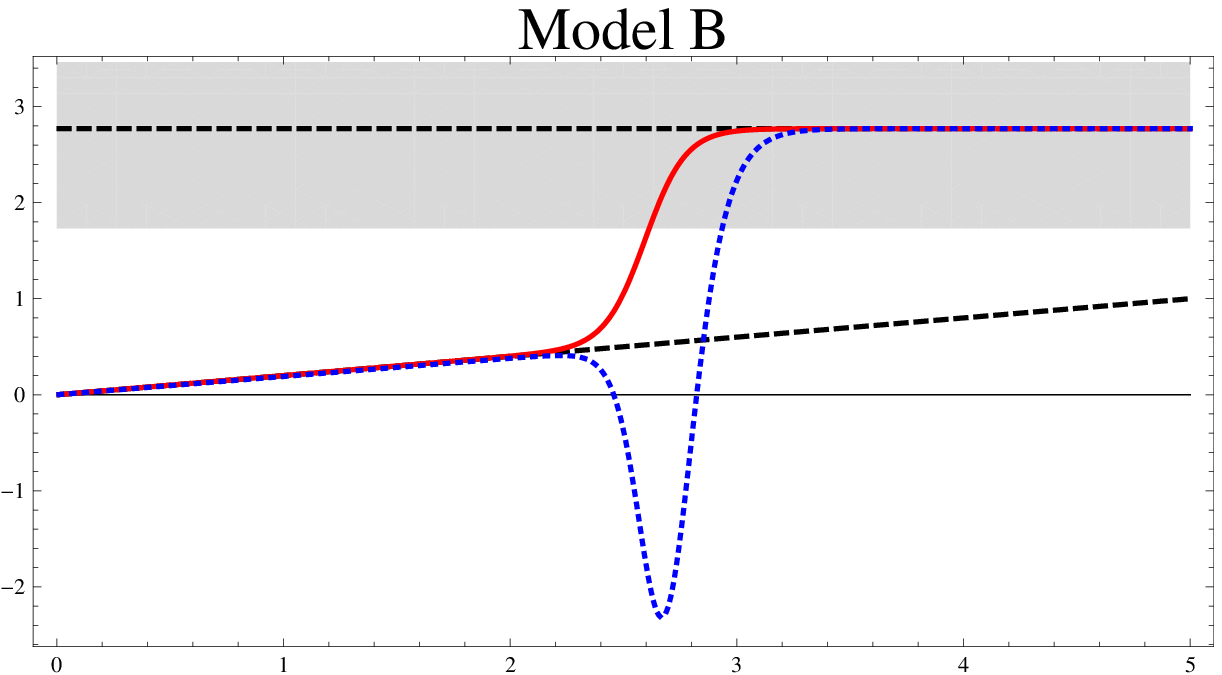} \\
(A)
    & 
      (B)
\end{tabular}
\caption{(A) $\frac{(d-1)}{2}\frac{V^\prime}{V}$ (blue dotted) and $-\beta(\phi)$ (continuous red) as a function of $\phi$ for the model A defined in Eq.~(\ref{eq:modelA}). The three regimes (deformation, condensation and confinement) are clearly distinguishable in $\beta(\phi)$. (B) The same as (A) for model B defined in Eq.~(\ref{eq:modelB}).}
\label{fig:modelAB}
\end{figure*}

Up to now we have presented all the background material on holographic RG flows. We are now ready to specify two concrete holographic models for CFTs with confining deformations to illustrate the physics of light dilatons. Model A corresponds to a non-zero value of the condensate $\langle \OO \rangle \ne 0$, and it is defined in terms of the scalar potential as
\begin{equation}
\frac{V^\prime(\phi)}{V(\phi)} = \Bigg\{
\begin{tabular}{cc} 
$\frac{2 \Delta_- \Delta_+ }{ d(d-1) } \phi$  \qquad \hspace{-0.15cm} {\rm  for} & $\quad \phi < \phi_{conf}$ \\
$2\nu \sqrt{\frac{d}{d-1}} \qquad$ {\rm  for}  & $\quad \phi > \phi_{conf}$ 
\end{tabular} \,. \label{eq:modelA}
\end{equation}
The behavior of the beta function is then obtained by solving Eq.~(\ref{eq:eqbeta}), and the integration constant $s$ is fixed as explained in Sec.~\ref{subsec:CFT_confining_def}. The resulting RG flow is displayed in Fig.~\ref{fig:modelAB}(A). In addition to the confinement scale $a_{conf}$, we observe the appearance of a condensation scale $a_{cond}$. The regime $\phi_{cond} < \phi < \phi_{conf}$ is a condensate-dominated region in which $\beta(\phi)$ is well approximated by a condensate type flow, i.e. $\beta_+ \simeq -\Delta_+(\phi - \phi_{cond})$.

It is also possible to design a model with no condensate, i.e.~$\langle \OO  \rangle = 0$, just by providing an analytical expression for $\beta(\phi)$, so that $\beta_\NPP(\phi)=0$ in Eq.~(\ref{eq:betaPNP}). We will refer it as model B, and it is defined as
\begin{equation}
\beta(\phi) = \beta_{\PP}(\phi) = \Bigg\{
\begin{tabular}{cc} 
$-\Delta_-\phi  \qquad\qquad$ \hspace{0.15cm}  {\rm  for} & $\quad \phi < \phi_{conf}$ \\
$-\nu\sqrt{d(d-1)} \qquad$  {\rm  for} & $\quad \phi > \phi_{conf}$
\end{tabular} \,. \label{eq:modelB} 
\end{equation}
The oscillating behavior of $\frac{V^\prime}{V}$ shown in Fig.~\ref{fig:modelAB}(B) is a manifestation that the potential needs to be carefully adjusted to avoid the formation of a condensate. As we will explain in Sec.~\ref{sec:spectra} both models lead to a light dilaton, and the main difference is the fine-tuning character of model B which is manifest in the precise value of the condensate.

\section{Spectra and mass formula: the dilaton}
\label{sec:spectra}

In order to compute the scalar spectrum and in particular its lightest mode, we have to consider a scalar perturbation of the background flow solution. The most general one is given by allowing
\begin{equation}
\phi=\phi_0(y)+\delta\phi
\qquad {\rm and} \qquad ds^2 = N^2 dy^2 + g_{\mu\nu}(dx^\mu+N^\mu dy)(dx^\nu+N^\nu dy) \,, \label{eq:metricscalar}
\end{equation}
with $N=1+\delta N$, $N_\mu=\partial_\mu \psi$, $g_{\mu\nu}=a^2(y)\,e^{2\xi}\,\eta_{\mu\nu}$.~\footnote{We refer the reader to Refs.~\cite{Maldacena:2002vr,Megias:2014iwa} for details on the computation.} In the unitary gauge, $\delta\phi=0$, one finds the following reduced action for $\xi$
\begin{equation}
S_{red} = {M^{d-1}\over 2} \int d^{d}x dy \;a^d {\beta^2}\; \left[{(\partial_\mu \xi)^2\over a^2} - (\partial_y \xi)^2 \right]  \label{redaction}  \,,
\end{equation}
in which we have used the background equations of motion, and integrated by parts a number of times. Then the spectrum can be obtained from the 2nd order differential equation
\begin{equation}
\frac{1}{a^d \beta^2} \partial_y \Big[ a^d \beta^2 \partial_y \xi_n\Big] + \frac{m_n^2\xi_n}{a^2} = 0 \,. \label{eq:eomxi}
\end{equation}
To get this equation we have considered a Fourier decomposition in the boundary coordinates $\xi = \xi_n e^{-i k_\mu x^\mu}$, with the mode masses $k_\mu k^\mu = -m_n^2$. Since $\xi + \delta\phi/\beta$ is a gauge invariant variable, one can translate from the $\xi$ variable into a $\delta\phi$ perturbation by adding a factor $\beta$ , i.e.~$\delta\phi = \beta \xi $. In the following we will work in the standard quantization, so that we will require that $\delta\phi$ contains only the sub-leading mode, i.e.~$\delta\phi \propto e^{-\Delta_+ y}$.~\footnote{A study in other quantizations will be presented in a forthcoming publication~\cite{Megias:2015inprogress}.} This, in addition to the requirement of regularity in the IR, leads to a discrete spectrum $m_n$. 

Firstly, we will make an analytical study of the fluctuations for massless and light dilatons, and then present the full numerical computation with Eq.~(\ref{eq:eomxi}).

\subsection{Massless dilaton}
\label{sec:massless_dilaton}

By setting $m_n=0$ in Eq.~(\ref{eq:eomxi}) one finds the following exact solution for the eigenvalue problem
\begin{equation}
\xi_{m^2=0} = c_1 + c_2 \int_0^z dz' \frac{1}{a^{d-1} \beta^2} \,.
\end{equation}
The mode $c_2$ diverges at the IR singularity, so that $c_2=0$. Note however that the solution $\xi_{m^2=0} = c_1$ leads to $\delta\phi = c_1 \beta \sim z^{\Delta_-}$, so that it does not satisfy the standard quantization prescription unless $c_1=0$. This is interpreted as the impossibility of getting a massless dilaton in CFT deformations, a feature already anticipated in Sec.~\ref{sec:SBCI_dilatons}.

\subsection{Light dilaton}
\label{sec:light_dilaton}

Let us now study under which circumstances a light dilaton can appear in the spectrum, which means the existence of a mode with mass $m_{\dil} \ll \Lambda_{\IR}$. This problem allows for an analytical treatment based on the method of matched asymptotic expansions~\cite{Hoyos:2013gma,Bednik:2013nxa,Megias:2014iwa}. Treating $m^2$ as a small parameter in Eq.~(\ref{eq:eomxi}), one can obtain the following UV and IR expansions of the solution
\begin{equation}
\xi_\dil^{\alpha} = \left(1- m^2 \,{\cal I}_\alpha +\dots\right) \left[c_1^\alpha + c_2^\alpha \int_{z_\alpha}^z dz' {1\over a^{d-1}\,\beta^2}\right]  \,, \qquad {\rm for} \qquad \alpha = {\rm UV}, {\rm IR} \,, \label{eq:xi_asymptotic}
\end{equation}
where $ z_\UV = 0$ and $z_\IR = z_s$. The integral operators act to the right, and they are defined as
\begin{equation}
{\cal I}_{\alpha} = \int_{z_\alpha}^z dz' \frac{1}{a^{d-1}\beta^2} \int_{z_\alpha}^{z'} dz'' a^{d-1} \beta^2 \,.
\end{equation}
Regularity in the IR and standard quantization select $c_2^\IR=0$ and $c_1^\UV = 0$ respectively. One expects the existence of an overlap region, $ \frac{z_s}{2} \lesssim z \ll z_s$, where $\xi_\dil^\UV(z) \approx \xi_\dil^\IR(z)$. Then one can impose the matching conditions
\begin{equation}
\xi_\dil^{\UV}(z_\ma)=\xi_\dil^{\IR}(z_\ma) \qquad {\rm and} \qquad {\xi_\dil^{\UV}}'(z_\ma)={\xi_\dil^{\IR}}'(z_\ma)  \,, \label{eq:matching_eq}
\end{equation}
at any point $z_\ma$ in this region. Finally from Eq.~(\ref{eq:matching_eq}) one can easily obtain the solution
\begin{equation}
\frac{c_2^\UV}{c_1^\IR} \simeq  - m_\dil^2 \, \int_{0}^{z_s} dz \, a^{d-1}\,\beta^2 \,, \qquad {\rm and } \qquad
m_\dil^{2} \simeq  
\left[ \int_{0}^{z_s} dz \; {1\over a^{d-1} \beta^2} \int_{z}^{z_s} dz'{a^{d-1} \beta^2}  \right]^{-1} \,.  \label{eq:mass_formula}
\end{equation}
In these expressions we have made the choice $z_\ma = 0$, which in fact it has been found to be a good approximation, see Ref.~\cite{Megias:2014iwa}. The mass formula given by Eq.~(\ref{eq:mass_formula}) constitutes one of the main results of this work. Note that in the absence of fixed points apart from the UV one, i.e. $\beta(\phi) \ne 0$, this formula predicts $m_\dil^2>0$, leading to stability of the theory.

\begin{figure*}[htb]
\begin{tabular}{cc}
\includegraphics[width=75mm]{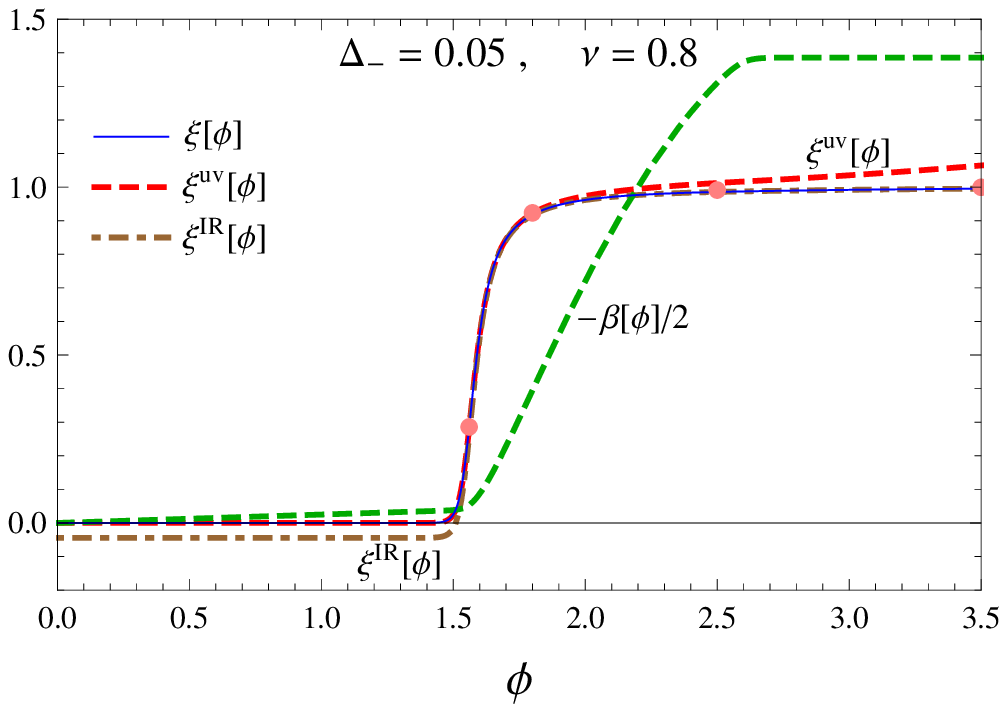} & 
\includegraphics[width=75mm]{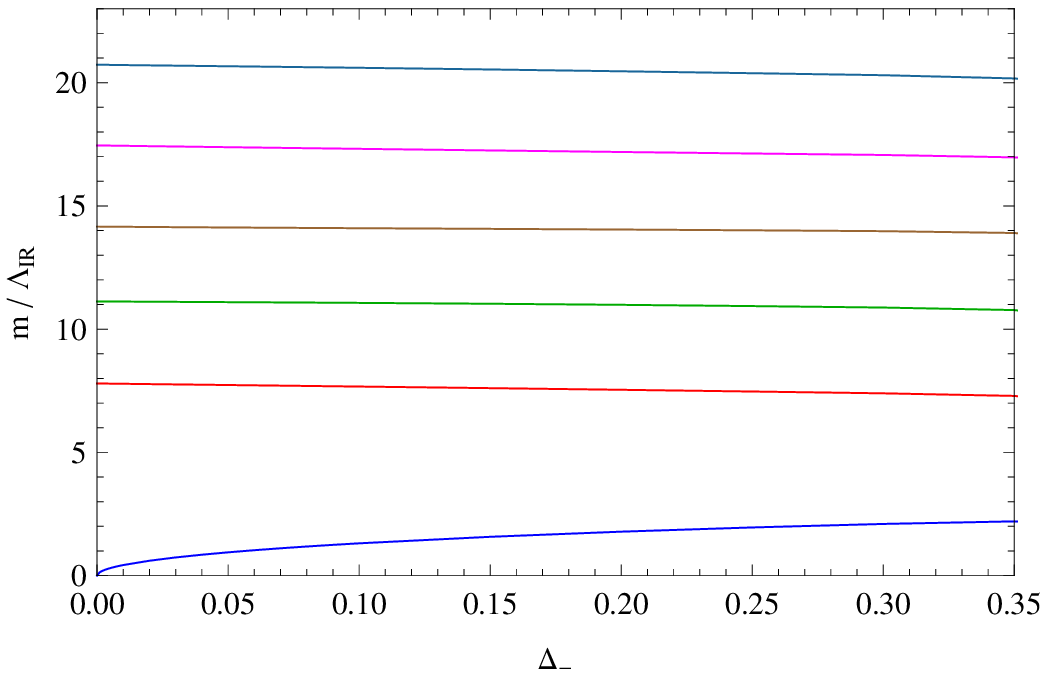} \\
(A)
    & 
      (B)
\end{tabular}
\caption{(A) Dilaton wave function $\xi$ as a function of $\phi$, obtained by numerically solving Eq.~(\ref{eq:eomxi}) (continuous line) and the UV- and IR-approximations of Eq.~(\ref{eq:xi_asymptotic}) (dashed lines). It is also displayed the background beta function. (B) Spectrum as a function of $\Delta_-$ from a numerical computation of Eq.~(\ref{eq:eomxi}). The first 6 modes are depicted. We have used in both figures the model~A defined in Eq.~(\ref{eq:modelA}). Similar results are obtained for model B defined in Eq.~(\ref{eq:modelB}), except for the dilaton mass scaling with $\Delta_-$.}
\label{fig:spectrum}
\end{figure*}

We show in Fig.~\ref{fig:spectrum} the wave function and the spectrum obtained with a numerical solution of the equation of motion for the fluctuation, Eq.~(\ref{eq:eomxi}). It is obvious from the figure that there is a light state, which in fact is well described by Eq.~(\ref{eq:mass_formula}). We find that the lightest mode --  the dilaton -- scales like $m_\dil^2/\Lambda_\IR^2 \sim \Delta_-$ and $\sim \Delta_-^2$ for model A and B respectively, so that it becomes light for nearly marginal deformations, i.e. $\Delta_- \ll 1$.

It is possible to compute the integral of Eq.~(\ref{eq:mass_formula}) analytically in a toy model which captures the main features of the flow. By extending the model B of Eq.~(\ref{eq:modelB}) with a condensation region, i.e. $\beta(\phi) = -\widetilde\Delta_+ (\phi-\phi_{cond}) - \Delta_- \phi_{cond}$ for $\phi_{cond} < \phi < \phi_{conf}$, one finds the following result for small~$\Delta_-$~\cite{Megias:2014iwa}
\begin{equation}
m_\dil^2 / \Lambda_\IR^2 \sim \beta_{cond}^{2-{d/\widetilde\Delta_+}} \,. \label{eq:m_betacond}
\end{equation}
The dilaton mass is suppressed so long as $\beta_{cond} \ll 1$ and $\widetilde\Delta_+ > \frac{d}{2}$. In particular, this model reproduces the suppression of $m_\dil$ for model A when $\widetilde\Delta_+ = \Delta_+$, and model B when $\widetilde\Delta_+ \gg 1$. Eq.~(\ref{eq:m_betacond}) illustrates the main result of this work, i.e. that the dilaton is light whenever i) the beta function is small in the IR (more precisely at the condensation scale) and ii) the rise in $\beta$ towards confinement $(\beta_{conf} \sim 1)$ is fast enough. The natural way that the beta function experiences a fast rise is by the condensation of $\langle \OO \rangle$ itself.

\section{Discussion and conclusions}
\label{sec:conclusions}

In this work we have studied the realization of a light dilaton by considering CFTs deformed by a single nearly-marginal scalar operator $\OO$. Such deformation induces generically a nonzero condensate $\langle \OO \rangle$, which we have called models of type~A. In addition, we have found a family of fine-tuned models, called models of type~B, with a vanishing value for the condensate $\langle \OO \rangle =0$. The dilaton mode is identifiable with the fluctuation of the condensate, regardless of whether $\langle  \OO \rangle$ vanishes or not. In both cases we find that the dilaton mass is controlled by the value of the beta function in the IR as conjectured in Ref.~\cite{CPR}, i.e. $m_\dil^2 \sim \beta_\IR$. More precisely, the dilaton is light when $\beta_\IR$ is small and the rise towards confinement is fast enough.

It has been found a rather simple mass formula for the dilaton mass involving an integral over the flow, cf. Eq.~(\ref{eq:mass_formula}), which can be used as a diagnostic for the presence of a light dilaton in the spectrum. This formula is explicitly positive, and this can be seen as a perturbative stability result. This allows to identify the regular RG flow as the ground state of the CFT deformation.

Finally we have studied the method to compute the condensate~$\langle \OO \rangle$ resulting from the CFT deformation. Holographically, this is based on the solution of a first-order differential equation for the beta function, supplemented with a regularity condition in the IR. This uniquely determines the condensate once the holographic model is specified.

There remain some open questions, in particular i) the generalization of these results to other quantizations like alternative and double trace deformations, and ii) the study of stability and the possible existence of tachyons in more general backgrounds. There are in the literature some computations of the effective potential in related theories, see e.g.~\cite{Kiritsis:2014xna}, and this can be used as well to shed some light on these and other issues. A study of these topics is currently in progress~\cite{Megias:2015inprogress}.

Concerning possible applications with a naturally light dilaton, one of the most obvious is adopting the light dilaton as the Higgs boson found at the LHC. Within the present mechanism this would allow to describe the light mass of the Higgs without fine-tunings. To test this possibility it should be studied the coupling of the dilaton with SM matter fields
\begin{equation}
{\cal L}_{\phi-T_{\mu\nu}} = \frac{1}{f_d} \phi \, T^\mu_\mu \,,
\end{equation}
and compare the strength of the couplings with the ones predicted for the Higgs by the~SM~\cite{Csaki:2007ns,Goldberger:2008zz}. It is important to stress that even if the dilaton were not a Higgs {\it imposter}, it could affect its phenomenology in such a way that it might be detected from deviations of the SM predictions at the LHC or future colliders. In any case, this could be the starting point for beyond the SM phenomenology: extra-dimensions and/or strongly coupled sector; and the dilaton would be naturally the lightest state of this sector of new physics. This study is currently in progress~\cite{Megias:2015inprogress2}.

Other target is to study the dilaton as a (light) dark matter candidate. In this case it could happen that the coupling of the dilaton to matter fields~\cite{Cembranos:2008gj} would require the assumption of additional symmetries in order to get stability at cosmological scales. Finally, it deserves to be investigated the CC problem in light with a mechanism similar to the one presented in this work, see e.g.~\cite{Bellazzini:2013fga}.

\ack
This work is based on Ref.~\cite{Megias:2014iwa}, co-authored with O.~Pujol\`as. We thank M.~Baggioli, B.~Bellazzini, A.~Pineda, J.~Serra, W.~Skiba, M.~Quir{\'o}s and especially A.~Pomarol for valuable discussions. We acknowledge support from the European Union under a Marie Curie Intra-European fellowship (FP7-PEOPLE-2013-IEF) with project number PIEF-GA-2013-623006.

\section*{References}

\bibliography{Megias}

\end{document}